\documentclass[preprint,
superscriptaddress,
amsmath,amssymb,aps,showkeys,showpacs,
twoside,final,secnumarabic,
nofootinbib]{revtex4-2}

\usepackage[paperwidth=205mm,paperheight=290mm,top=17mm,bottom=25mm,
inner=17mm,outer=17mm,
twoside]{geometry}

\usepackage{cmap} 
\usepackage[T1,T2A]{fontenc}
\usepackage[utf8]{inputenc}
\usepackage[russian,english]{babel}
\usepackage{color}
\usepackage{graphicx}
\usepackage{dcolumn}
\usepackage{bm} 
\usepackage[unicode=true,colorlinks=true,linkcolor=magenta, urlcolor=blue, citecolor = blue,breaklinks]{hyperref}
\usepackage{multirow}
\usepackage{url}
\usepackage{breakurl}
\DeclareGraphicsExtensions{.eps}

\newcount\issue
\newcount\Vol
\newcount\numb
\usepackage{fancyhdr} 
\def\Vol{\textbf{80}}
\def\numb{x}
\begin{document}

\title{GALAXIES\\[20pt]
Star formation in low-density regions\\ of galactic disks} 

\def\addressa{Sternberg Astronomical Institute, Moscow State University, Universitetsky pr., 13, Moscow, 119234, Russia}

\author{\firstname{A.V.}~\surname{Zasov}}
\email[E-mail: ]{zasov@sai.msu.ru }
\affiliation{\addressa}
\author{\firstname{N.A.}~\surname{Zaitseva}}
\affiliation{\addressa}
 \author{\firstname{A.S.}~\surname{Saburova}}
\affiliation{\addressa}

\received{xx.xx.2025}
\revised{xx.xx.2025}
\accepted{xx.xx.2025}

\begin{abstract}
We argue that star formation in the disks of low-surface-brightness (LSB) galaxies shares a similar nature with that occurring in the far outer regions of normal-brightness spiral galaxies, such as those with the extended ultraviolet (XUV) disks. In both cases, stars are born in gravitationally stable disks with an extremely low average gas density (on kiloparsec scales), and the efficiency of this process depends on a disk brightness in a similar way. Processes which can stimulate star formation under these conditions are shortly discussed. 
\end{abstract}

\pacs{Suggested PACS}\par
\keywords{Suggested keywords   \\[5pt]}

\maketitle
\thispagestyle{fancy}


\section{Introduction}\label{intro}

Star formation rate (SFR) in galaxy disks is primarily determined by the local values of gas density and gas pressure in the disk plane.    
However in some cases we observe signs of star formation which is taking place despite very low gas density, raising the question of the mechanisms that can stimulate and sustain this process. These are low-surface-brightness galaxies (LSB-galaxies), which have a specific formation history, and extended far-fetched disks, often found through their UV radiation (XUV-disks) in some normal-brightness galaxies with active star formation (hereafter referred to as HSB galaxies).  

\section{\label{sec:level1}LSB vs XUV disks}

LSB-galaxies are a rare type of disk galaxies where even the central (or extrapolated to the center) surface brightness does not exceed the optical brightness of the dark night sky and, in many cases, is significantly below it. At the same time, such galaxies may have a bulge that is normal in terms of stellar composition. They are characterized by a large mass of atomic hydrogen (HI) for a given disk luminosity, and, as a rule, by a very low content of molecular gas relative to HI, evidently due to low gas density and low metal abundance.  
Due to the sparseness of a stellar disk, the mass of dark halo in LSB-galaxies significantly exceeds the mass of their stellar components. At the same time, the integral gas (HI) masses and the rotation velocities of LSB-galaxies are typical for spiral galaxies with high surface brightness. Despite the extremely low density of their  disks, weak star formation continues in LSB-galaxies, manifesting itself both in the presence of emission regions and in the UV disk radiation. Structural details are often observed there, such as bars, fragments of spirals, or ring arcs. 
Possible ways of formation of low brightness disks in LSB-galaxies remain poorly understood (see, for example, the discussion in \cite{1,2}).

Another example of star formation in gas low-density conditions is outermost regions of normal-brightness spiral galaxies. As a consequence of the decrease in disk density and the preservation of sufficiently high observed HI velocity dispersion, a gaseous disk at the periphery flares, its half-thickness should reach  hundreds of parsecs and even more, so its volume gas density turns out to be extremely low (see, for example, \cite{3,4}). Nevertheless, these sparse disks contain newly formed stars. Observations made by the GALEX UV observatory have shown that in significant part of spiral galaxies, weak UV radiation associated with young stars extends far beyond the optical radius $R_{25}$, corresponding to a brightness of $25^m/arcsec^2$. 

\begin{figure}[b]
\includegraphics{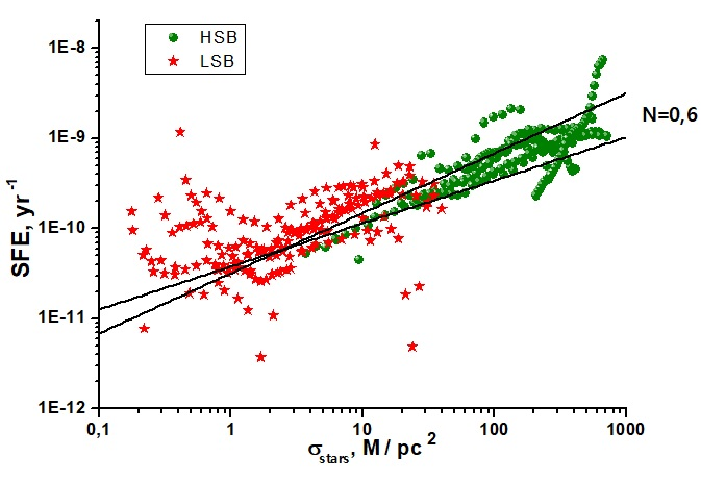}
\caption{\label{fig:1} A variation of  star formation efficiency with respect to a surface density of stellar disks for LSB- (red asterisks) and HSB- (green) galaxies without the extended disks (see \cite{6}).}
\end{figure}

To compare SFR observed at various gas densities in the disks, the volume gas density ($\rho_g$) was calculated in the plane of the HSB- and LSB-galaxies in (\cite{6}) and compared with the measured radial profiles of star formation rates (taken from the GALEX data). It was shown that these galaxies form a united sequence in the $SFR(\rho_g)$-plane, going from high gas density down to  $(\rho_g)\sim 10^{-26}g/cm^3$. A similar sequence, common for LSB and HSB-galaxies, appears if one compares the star formation efficiency $SFE = SFR/M_{HI}$ (star formation rate per unit mass of HI) with the surface density of their stellar disks (Fig.~\ref{fig:1}).

Later, a similar dependence was found in \cite{7}. The slope of the dependence is close to 0.5, which is expected in the model where star formation has a quasi-stationary character, while SFR is determined by the gas pressure in the plane of the equilibrium gas-star disk (\cite{9}).

\begin{figure}[b]
\includegraphics[width=0.8\textwidth]{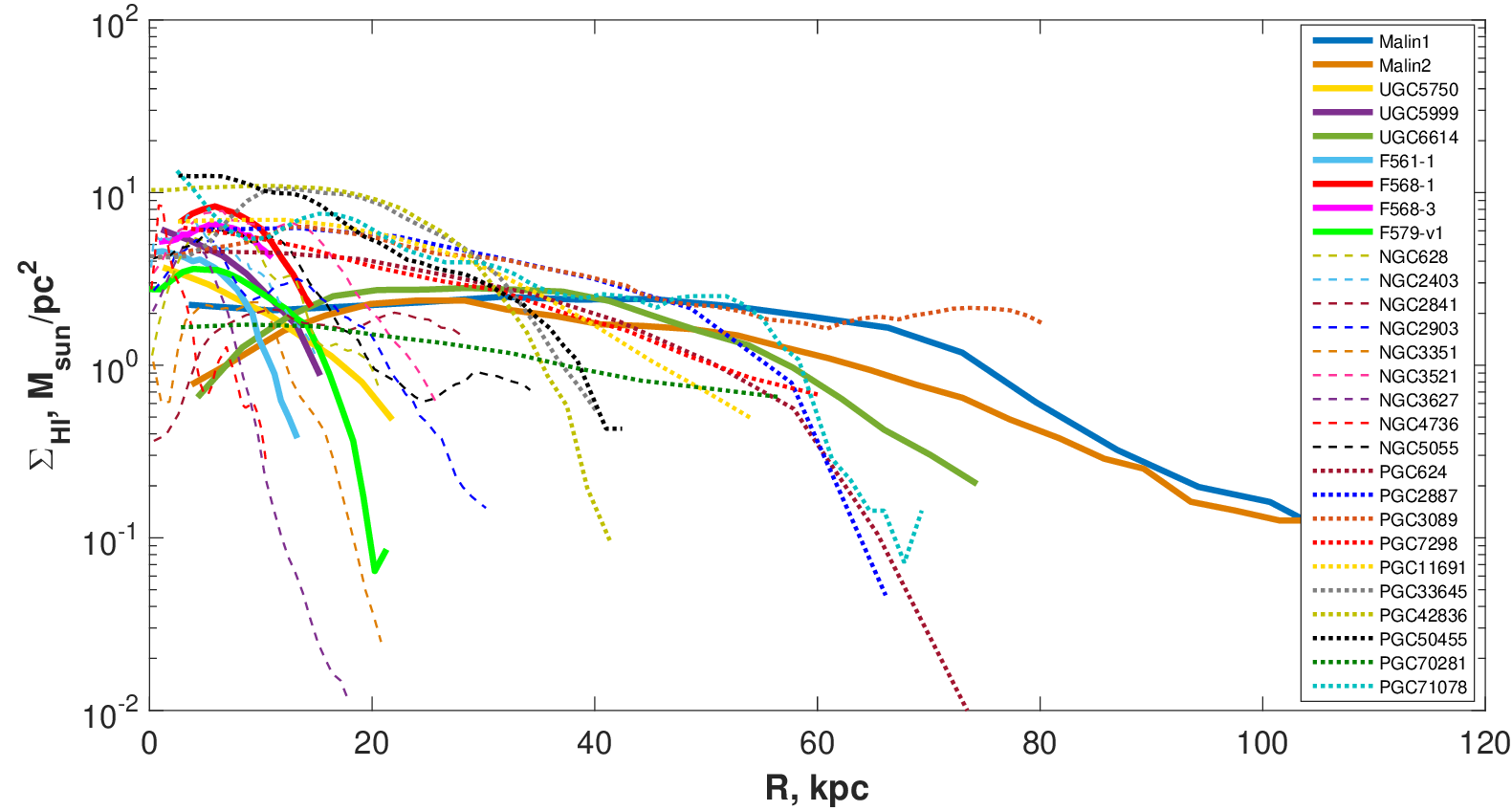}
\caption{\label{fig:2} Radial profiles of HI surface density for the galaxies considered. All three subsamples of galaxies are presented here: bold lines represent LSB-galaxies, dashed lines are for gas-rich spirals (HIR-galaxies), and dotted lines represent THINGS galaxies.}
\end{figure}

In Fig.~\ref{fig:2} the radial distribution of HI density is shown (the data were taken from the literature) for LSB-galaxies (thick lines) and those HSB-galaxies where HI is traced beyond the bright spiral structure. The latter are galaxies from the THINGS sample (\cite{10}) with the available extended $SFR(R)$ profiles, and galaxies with abnormally high HI content (\cite{11}), in which the gas mass is several times higher than the average one for galaxies of the same luminosity $L_K$. Being HSB-galaxies, they may nevertheless possess extended disks of low brightness beyond the optical radius $R_{25}$, that makes them look like disks of LSB-galaxies. At the same time, these galaxies do not differ from spiral galaxies with a lower gas content in terms of metallicity gradient, making the external accretion improbable cause of their high HI mass (\cite{12}). As can be seen in Fig.~\ref{fig:2}, the LSB-galaxies under consideration (bold lines) do not differ in terms of abnormal gas surface density values: the radial density profiles for giant LSB-galaxies are similar to those for HI-rich spiral galaxies, and smaller LSB-galaxies can be traced in a similar way as normal THINGS spiral galaxies.

\begin{figure}[b]
\includegraphics[width=0.8\textwidth]{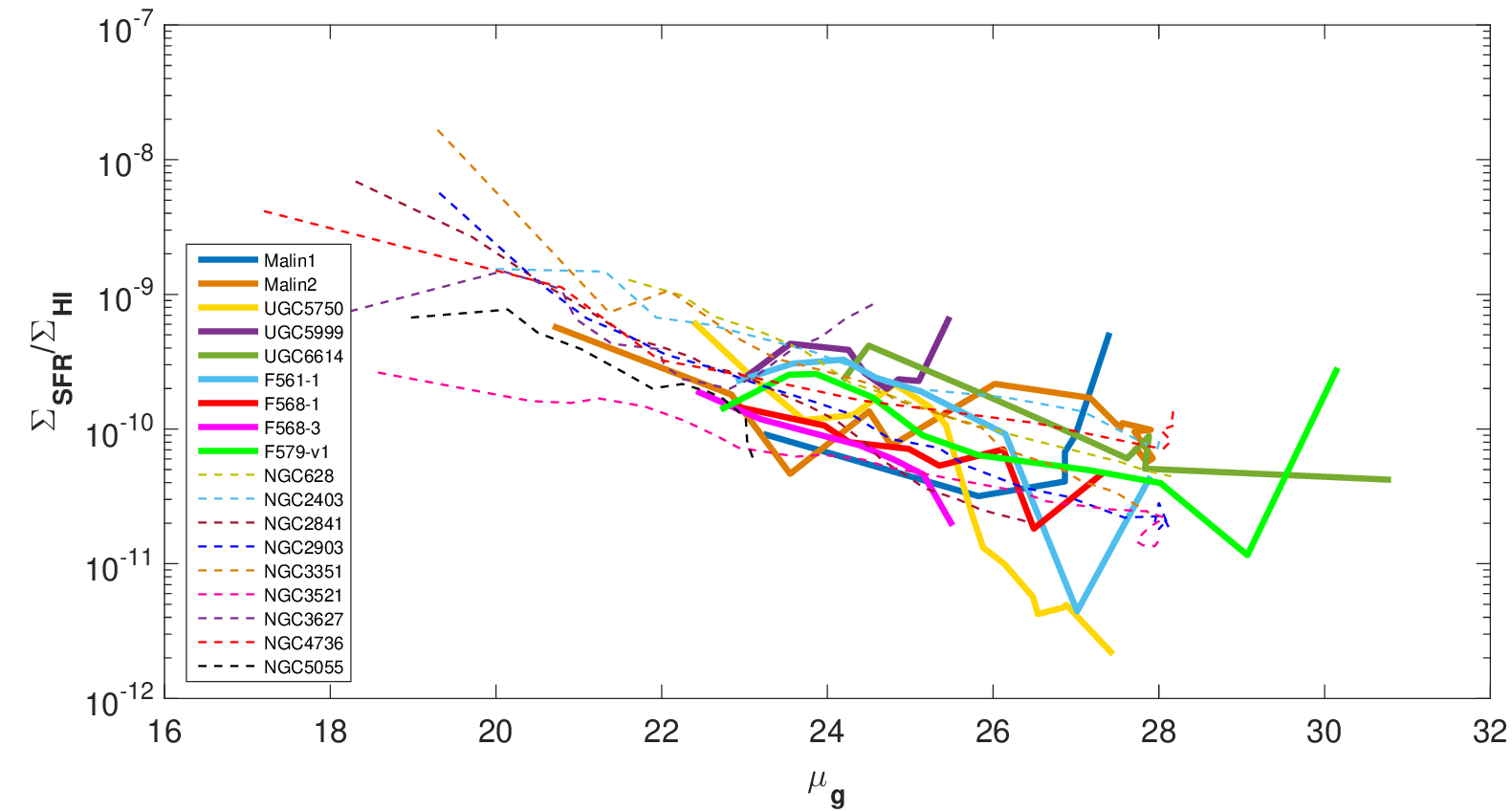}
\caption{\label{fig:3} Comparison of star formation efficiency for LSB- and THINGS galaxies at the same surface brightness $\mu_g$ in $g$-band. The designations are similar to Fig.~\ref{fig:2}.}
\end{figure}

The distinctive feature of LSB-galaxies compared to ordinary spiral galaxies is not the content and distribution of gas or the maximal rotation velocity, but the low brightness (density) of their stellar disks. At the same time, star formation rate  (SFE) per unit area of a disk, which is assumed to be proportional to the surface brightness in the UV range, corresponds well to the same HI surface density as for galaxies of normal brightness. This is illustrated in Fig.~\ref{fig:3}, where SFE is compared with the local brightness $\mu_g$ for the disks of LSB- and HSB-galaxies. Disk surface brightness here characterizes a surface density of stellar disk, although it cannot be excluded that the mass-to-light ratio in the rarefied disks may differ from that in the dense disk regions. As it can be seen from Fig.~\ref{fig:3}, for regions with similar brightness (density), there is no systematic difference between LSB- and HSB-galaxies in terms of star formation rate and efficiency. The existence of relationship between SFE and the density (brightness) of stellar disks suggests that even at low disk densities, star formation is not random or sporadic process, as might be expected, for example, if it takes place in local regions of external gas accretion (note that in the most remote regions of disks the observed star formation sites are generally scattered across a disk, and, apparently, no longer associated with the stellar disk components). 

\begin{figure}[b]
\includegraphics[width=0.8\textwidth]{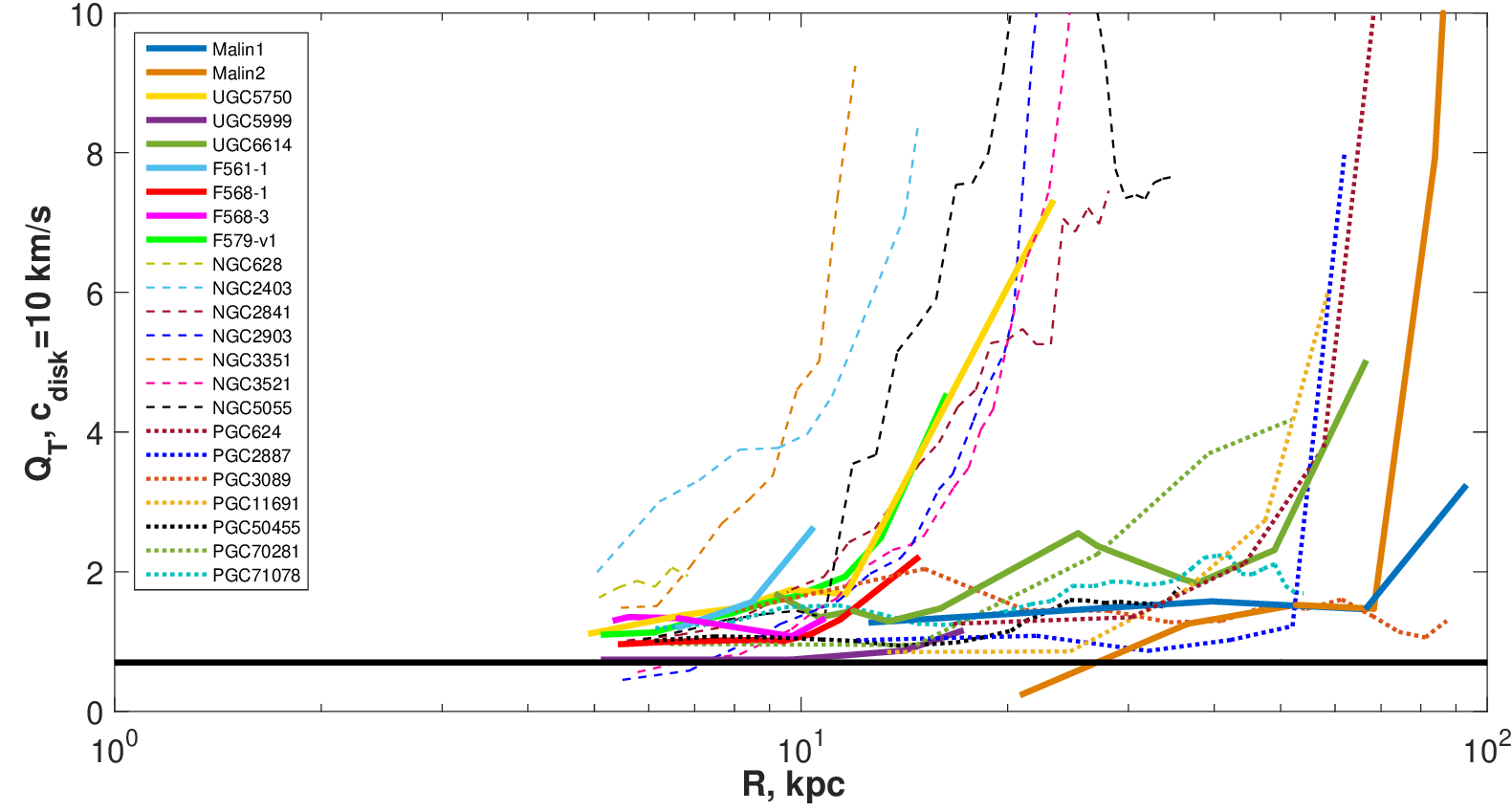}
\caption{\label{fig:4} Radial distribution of the Toomre stability parameter for LSB- and HSB-galaxies. The designations are the same as in Fig.~\ref{fig:2}. The black horizontal line marks $Q_T\sim0.7$.}
\end{figure}

A similarity between the LSB-disks and the outer rarefied  regions of normal-brightness galaxies is also illustrated in Fig.~\ref{fig:4}, which shows the radial variation of the gas-star disk stability parameter $Q_T$ (Toomre' parameter). Here it was calculated for the radial stellar and gaseous velocity dispersion (isotropic sound velocity), which assumed to be 10 km/s (for HI, this value follows from observations of galaxies outer regions with low star formation, although in some cases it turns out to be slightly lower). The horizontal line $Q_T\approx 0.7$ corresponds to the threshold value for the gravitational stability of a disk of finite thickness, which takes into account the velocity dispersion along the z-axis (according to \cite{14}).

Despite the rather rough estimates made, they confirm the disks gravitational stability, suggesting at the same time that the disks of LSB-galaxies are not too overheated in spite of their low density (evidently due to the low angular velocity). For most of the LSB galaxies considered, the parameter $Q_T$ which is proportional to the ratio of angular velocity to surface density of the disk varies slightly with $R$, remaining in most cases within the range of 1--2.5 (at the accepted value of HI velocity dispersion) over the large range of distances $R$ from the center. It allows to conclude that the disk density in the considered galaxies correlates with its angular velocity. Note that the latter depends primarily on the mass distribution (the radial density scale) of a dark halo which dominates in mass. This connection between the disk densities and halo properties allows us to propose that the low brightness (density) of disks in most cases is the result of internal processes taking place during the formation of LSB-disks in the gravitational field of the halo.

The question remains what stimulates the emergence of dense star forming regions in the rarefied disks of LSB- and HSB-galaxies. Obviously, different scenarios may play out in different cases. The role of external influences in stimulating star formation is well known. These include tidal effects, the accretion of gas, and the infall of satellites onto a disk, or the blowing of the galaxy gas by the intergalactic medium as it moves in a cluster. However, these types of influences do not explain neither the relationship between star formation efficiency and disk density, nor the approximate constancy of $Q_T$ over a wide range of radial distances in some LSB- and HSB-galaxies (excluding the most distant disk regions where $Q_T$ is always $\gg1$). Therefore, for those galaxies, which are non-members of clusters or rich groups and do not have close neighbors of comparable mass, the internal processes capable of sustaining weak star formation are most likely to account for the observed star formation in dynamically stable rarefied disks. These may include such events as the gas compression in spiral density waves (even low-contrast ones) in a sparse disk, induced, for example, by a triaxial dark matter halo (see, for example, \cite{15}), rare supernova explosions, which generate a compression wave in the gas medium and may support a relatively high HI velocity dispersion, as well as a passage of objects belonging to the halo population of galaxies (low-mass dwarf galaxies, HI clouds, or dark sub-halos) crossing a disk.

\section{\label{sec:level1}Main conclusions}

In most cases, the disks of LSB-galaxies can be regarded as systems with high gas content and initially low star formation efficiency, similar to that observed in normal-brightness galaxies far off the center. The relationship between star formation rates (per unit area of the disk) and the brightness (or density) of stellar disks, common for LSB- and HSB-galaxies, suggests that star formation in low-density regions in both cases  is primarily stimulated by internal processes within galaxies rather than by external influences on their gaseous disks. This agrees with the conclusion that the disks of LSB-galaxies do not appear to be strongly dynamically overheated over a significant range of radial distances.

\section*{FUNDING}
This project is supported by the Russian Science Foundation (RSCF) grant No. 23-12-00146.

\section*{CONFLICT OF INTEREST}

The author of this work declares that she has no conflicts of interest.



\begin{thebibliography}{}

\bibitem{1}
A.S. Saburova, I.V. Chilingarian, A.V.Kasparova, O.K. Sil'chenko, K.A. Grishin, I.Yu. Katkov and R.I. Uklein,
Monthly Notices of the Royal Astronomical Society, \textbf{503}, 830-849 (2021).
https://doi.org/10.1093/mnras/stab374 

\bibitem{2}
A.S. Saburova, I.V. Chilingarian, A. Kulier, G. Galaz, K.A. Grishin, A.V.Kasparova, V. Toptun and I.Yu. Katkov
Monthly Notices of the Royal Astronomical Society, \textbf{520}, L85-L90 (2023).
https://doi.org/10.1093/mnrasl/slad005 

\bibitem{3}
J.C. O'Brien, K.C. Freeman, P.C. van der Kruit
Astronomy and Astrophysics, \textbf{515}, id.A61 (2010).
https://doi.org/10.1051/0004-6361/200912566 

\bibitem{4}
С. Bacchini, F. Fraternali, G. Iorio and G. Pezzulli
Astronomy and Astrophysics, \textbf{622}, id.A64 (2019).
https://doi.org/10.1051/0004-6361/201834382 

%

\bibitem{6}
O.V. Abramova and A.V. Zasov
Astronomy Letters, \textbf{38}, 755-763 (2012).
https://doi.org/10.1134/S1063773712120018 

\bibitem{7}
K. Du, Y. Shi, Z.-Y. Zhang, Q. Gu, T. Wang, J. Wang, X. Li and S. Zhai
Monthly Notices of the Royal Astronomical Society, \textbf{518}, 4024-4037 (2023).
https://doi.org/10.1093/mnras/stac3341 

%

\bibitem{9}
E.C. Ostriker and C.-G. Kim
The Astrophysical Journal, \textbf{936}, id.137 (2022).
https://doi.org/10.3847/1538-4357/ac7de2 

\bibitem{10}
F. Walter, E. Brinks, W.J.G. de Blok, F. Bigiel, R.C. Jr. Kennicutt, Robert C., M.D. Thornley and A. Leroy
The Astronomical Journal, \textbf{136}, 2563-2647 (2008).
https://doi.org/10.1088/0004-6256/136/6/2563 

\bibitem{11}
K.A. Lutz, V.A. Kilborn, B.S. Koribalski, B. Catinella, G.I.G. Józsa, O.I. Wong, A.R.H. Stevens, D. Obreschkow, and H. Dénes
Monthly Notices of the Royal Astronomical Society, \textbf{476}, 3744-3780 (2018).
https://doi.org/10.1093/mnras/sty387 

\bibitem{12}
K.A. Lutz, V. Kilborn, B. Catinella, L. Cortese, T.H. Brownand B. Koribalski
Astronomy\& Astrophysics, \textbf{635}, id.A69 (2020).
https://doi.org/10.1093/mnras/sty387 

\bibitem{13}
T.K. Wyder, D.C. Martin, T.A. Barlow, K. Foster, P.G. Friedman, P. Morrissey, S.G. Neff, J.D. Neill, D. Schiminovich, M. Seibert, L. Bianchi, J. Donas, T.M. Heckman, Y.-W. Lee, B.F. Madore, B. Milliard, R.M. Rich, A.S. Szalayand S.K. Yi
The Astrophysical Journal, \textbf{696}, 1834-1853 (2009).
https://doi.org/10.1088/0004-637X/696/2/1834 

\bibitem{14}
A.B. Romeo and J. Wiegert
Monthly Notices of the Royal Astronomical Society, \textbf{416}, 1191-1196 (2011).
https://doi.org/10.1111/j.1365-2966.2011.19120.x 

\bibitem{15}
M.Butenko, A. Khoperskov and S. Khoperskov
Baltic Astronomy, \textbf{24}, 119-125 (2015).
https://doi.org/10.1515/astro-2017-0210 


\end{thebibliography}
\end{document}